\documentclass[amsmath,amssymb,pra,aps,showpacs,superscriptaddress]{revtex4-1}
\usepackage{amsmath,amsfonts,amssymb,amsthm,graphics,graphicx,epsfig}
\usepackage[colorlinks=true,citecolor=blue,linkcolor=blue,urlcolor=blue]{hyperref}
\usepackage[usenames]{color}
\usepackage{graphicx}
\usepackage{subfigure}
\usepackage{amsmath}
\usepackage{epsfig}
\usepackage{dcolumn}
\usepackage{bm}
\usepackage{color}
\usepackage{epstopdf}
\usepackage{amssymb}
\usepackage{amstext}
\usepackage{latexsym}
\usepackage{hyperref}
\usepackage{amsfonts}
\usepackage{psfrag}
\usepackage{xcolor}
\usepackage[normalem]{ulem}
\usepackage{dsfont}
\usepackage{txfonts}
\newcommand{\ket}[1]{\vert #1 \rangle}

\newcommand{\vacuum}{0}

\begin{document}

\title{Quantum detection of wormholes}

\author{Carlos Sab\'{i}n}  
\address{Instituto de F\'isica Fundamental, CSIC, Serrano 113-bis 28006 Madrid, Spain}
\email{Correspondence to: csl@iff.csic.es}
\begin{abstract}
We show how to use quantum metrology to detect a wormhole. A coherent state of the electromagnetic field experiences a phase shift with a slight dependence on the throat radius of a possible distant wormhole. We show that this tiny correction is, in principle, detectable by homodyne measurements after long propagation lengths for a wide range of throat radii and distances to the wormhole, even if the detection takes place very far away from the throat, where the spacetime is very close to a flat geometry. We use realistic parameters from state-of-the-art long-baseline laser interferometry, both Earth-based and space-borne. The scheme is, in principle, robust to optical losses and initial mixedness.
\end{abstract}

\maketitle

We have not observed any wormhole in our Universe, although observational-based bounds on their abundance have been established \cite{search}. The motivation of the search of these objects is twofold. On one hand, the theoretical implications of the existence of topological spacetime shortcuts would entail a challenge to our understanding of deep physical principles such as causality \cite{morristhorne,morristhorne2,hawking,deutsch}. On the other hand, typical phenomena attributed to black holes can be mimicked by wormholes. Therefore, if wormholes exist the identity of the objects in the center of the galaxies might be questioned \cite{combi} as well as the origin of the already observed gravitational waves \cite{gravastar,konoplya}. For these reasons, there is a renewed interest in the characterization of wormholes \cite{geodesics, taylor,mapping} and in their detection by classical means such as gravitational lensing \cite{lensing, lensing2}, among others \cite{shadows}.

Quantum metrology aims at providing enhancements to the measurements realised by classical means, by exploiting quantum properties such as squeezing and entanglement. This approach has already proven useful in a wide range of physical problems, from timekeeping \cite{clocks} to gravitational wave astronomy \cite{qgw}. While the successful observations of gravitational waves by advanced LIGO still did not leverage the benefits of squeezing, the technology is ready to be implemented in further upgrades, which is expected to significantly increase the detectors sensitivity \cite{future}. However, advanced LIGO has already been able to detect the effect of tiny spacetime oscillations on the phase of a coherent state of a laser containing a large number of photons. In the near future, several big scientific projects will be launched to space, including quantum metrology experiments and a long-baseline laser interferometer for gravitational wave detection -- LISA \cite{lisa}.  Inspired by these impressive technological developments, it is natural to ask whether it is possible to use similar technology for the detection of another deviations from a flat-spacetime geometry.
\begin{figure}[t!] 
\includegraphics[width=0.95\linewidth]{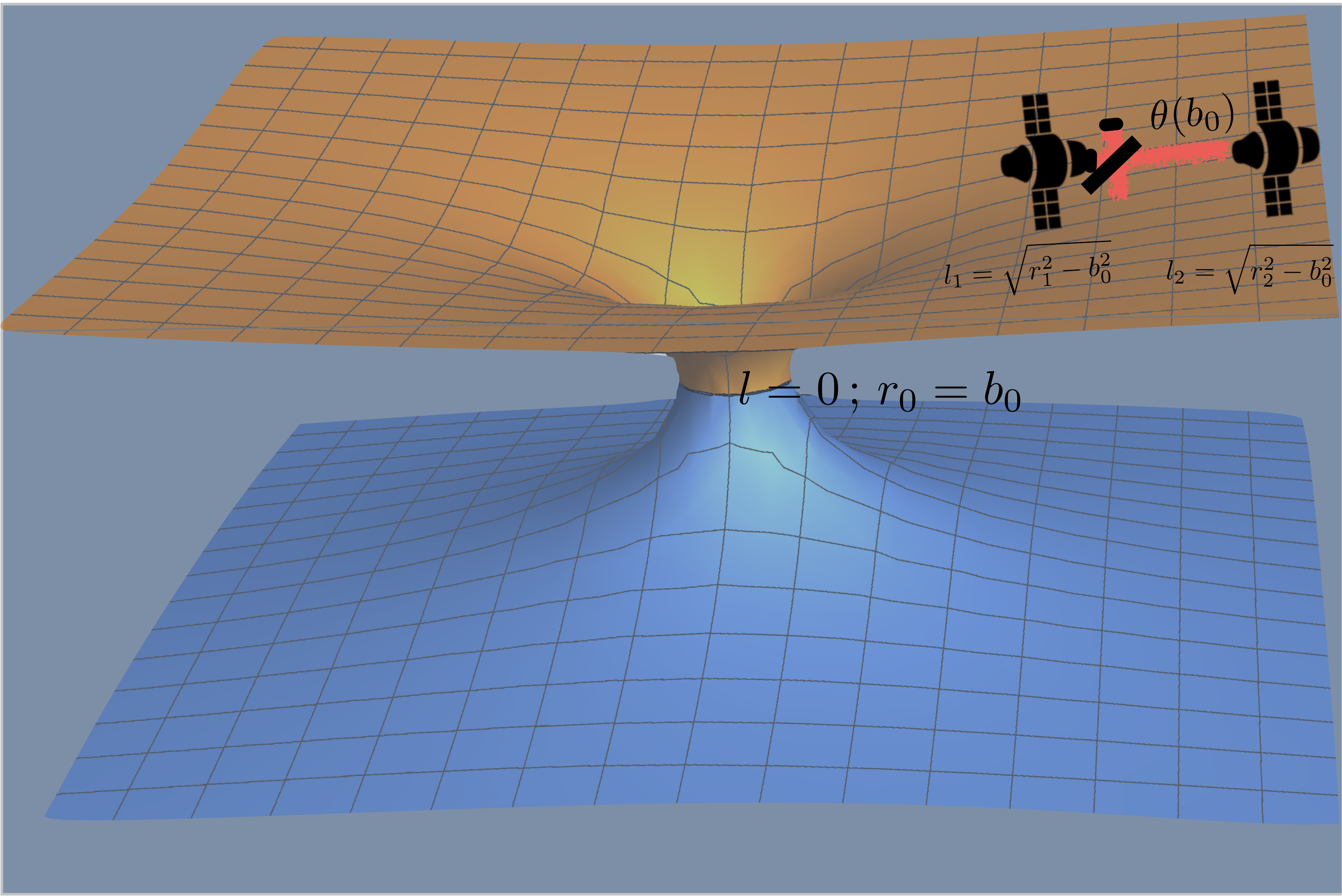}
 \caption{Sketch of the idea. A coherent single-mode state of a laser propagates along a large distance $l_2-l_1$, e. g. between two satellites. If the spacetime is totally flat the distance is just $r_2-r_1$. However, if the spacetime contains a distant wormhole at $l=0$ with a throat radius $b_0$, the traveled distance will be slightly different even if the spacetime is almost completely flat in that region, namely $r_2,r_1>>b_0$. This small correction  of the distance generates a slight phase shift, making the phase $\theta$ of the coherent state dependent on $b_0$. In principle, $b_0$ could be estimated by means of homodyne detection. Note that the length scale of the figure is not realistic.}
\label{fig1}
\end{figure}

In this work, we propose to use a quantum measurement scheme for the detection and characterization of wormholes (see Fig.\ref{fig1}. We show that a single-mode state of the electromagnetic field propagating in space will undergo a phase shift which would carry a slight dependence on the radius of the throat of a distant wormhole, in the case that the spacetime contains one of them. While this is of course a tiny correction with respect to the flat-spacetime case, we show that after large propagation lengths a good sensitivity is, in principle, within experimental reach by using parameters of modern long-baseline laser interferometers, even if propagation takes place very far away from the wormhole -- in a quasiflat spacetime geometry. We consider realistic initial  states and measurement protocols,finding that coherent states with large number of photons and homodyne detection respectively, are convenient choices. Indeed, for certain values of the phase shift, the homodyne measurement scheme achieves the ultimate quantum bound in the case of a coherent state. We show that the sensitivity is proportional to the ratio between the radius of the wormhole throat and the distance to the wormhole -- which is assumed to be very large -- providing a wide parameter window where the scheme is applicable. Furthermore, our results are, in principle, highly robust to the presence of optical losses and initial mixedness. Our aim is, of course, not to propose a detailed experimental setup, but to show  a new idea  that is, in principle, possible and might inspire further investigations, which would ascertain its actual feasibility.
\section*{Methods}
We start by considering a single-mode state of the electromagnetic field. By now, we will assume that the initial state is pure and we will discuss later the effects of temperature. Therefore, the initial state is characterized by:
\begin{equation}
    \label{eq:state}
    \ket{\phi_0}=D(\alpha)S(r)\ket\vacuum,
\end{equation}
where $S(r)=\exp[(r/2)(a^2-a^\dagger{}^2)]$ is a squeezing operator with real squeezing parameter $r$ and $D(\alpha)=\exp\left(\alpha a^\dagger-\alpha^*a\right)$ is a displacement operator with parameter ~$\alpha$. Here, $a$ and $a^{\dagger}$ are the standard annihilation and creation operators of the single mode.

Now we consider that the initial state in Eq. (\ref{eq:state}) undergoes unitary evolution characterized by the operator
$U(\theta)=\exp\left(-i\theta a^\dagger a\right)$, which depends on the phase $\theta$. Thus the resulting state is
\begin{equation} 
\ket{\phi_\theta}=U(\theta)D(\alpha)S(r)\ket\vacuum,
\end{equation}
 which is now $\theta$-dependent.
The phase shift $\theta$ can be estimated by realising measurements on the state $\ket{\phi_\theta}$. The ultimate quantum bound on the sensitivity $\Delta\theta$ of the state with respect to the parameter $\theta$ is given by the quantum Cramer-Rao bound \cite{braunsteincaves}:
\begin{equation}\label{eq:qcr}
\Delta\theta\geq\frac{1}{\sqrt{H(\theta)}},
\end{equation}
where $H(\theta)$  is the quantum Fisher information (QFI) of the state. Therefore, maximizing the QFI amounts to find the optimal bound for a given state. 
Indeed, it is well-known that it always exists an optimal measurement strategy which saturates the inequality in Eq. (\ref{eq:qcr}). In other words, the classical Fisher information (FI) of the optimal measurement matches the QFI. However, the optimal measurement might not be experimentally feasible in general.

For the single-mode pure state $\ket{\phi_{\theta}}$ the QFI is given by the variance of the generator of the phase, in this case the number operator \cite{monras,olivares}:
\begin{eqnarray}
    H(\theta)&=&4\langle\Delta a^\dagger a\rangle^2=\\
    \nonumber
    &=&4\left[|\alpha|^2\left(\cosh r-\sinh
    r\right)^2+2\sinh^2r\cosh^2r\right].
\end{eqnarray}
The average number of photons  of the state is
\begin{equation}
    \langle n\rangle=\langle a^\dagger
    a\rangle=|\alpha|^2+\sinh^2r.
\end{equation}
It is well-known \cite{monras, olivares} that for a fixed number of average photons the choice which maximizes the QFI is the squeezed vacuum $\alpha=0$, which attains a $\langle n\rangle^2$-scaling -- Heisenberg limit -- in contrast with the $\langle n\rangle$-scaling of the coherent state $r=0$ -- shot-noise limit. However, in practice it is very hard to experimentally achieve a purely squeezed vacuum with a number of photons high enough to take advantage of the above scaling. It is more feasible to create a coherent state with a large photon number, which in turn possesses a larger QFI than the squeezed vacuum state, when experimentally feasible squeezing parameters and photon numbers are  considered. Moreover, coherent states are more robust to typical imperfections, such as optical losses or thermal noise. Therefore, in the following we will consider coherent states $r=0$ and the QFI will be: 
\begin{equation}\label{eq:qfish}
    H(\theta)=4|\alpha|^2= 4 \langle n\rangle.
\end{equation}
Notice that the choice of the coherent state as a probe does not prevent us from using the benefits of squeezing or entanglement at a later stage of the measurement protocol. This is the same approach as in GEO 600 --which will soon be applied as well in advanced LIGO--, where a coherent state with a large number of photons is then mixed with a squeezed vacuum in a beam splitter. Indeed, this strategy has been shown to be very close to the optimal one -- which involves an entangled NOON state of large N-- in the experimentally relevant regime of optical losses and large number of photons \cite{rafalligo}. Furthermore, in the case of coherent states we know that the experimentally standard homodyne measurements are optimal. Indeed, the FI $F(\theta)$ associated to the homodyne measurement of the quadrature $\ket{p}$ is \cite{aspachs}:
\begin{equation}\label{eq:fish}
F(\theta)=4|\alpha|^2\cos^2(\theta).
\end{equation}
Therefore, homodyne measurements saturate the Cramer-Rao bound and provide the optimal sensitivity for particular values of the phase $\theta=m\pi,\,\, m=0,1,2,...$. In the following, we will restrict ourselves to coherent states, homodyne measurements and $\theta=m\pi$.

If the coherent state propagates between two points separated by a distance $L$ in flat spacetime, then the phase shift is simply $\theta_f=\omega t=\omega L/c$. Therefore, we can choose $L$ in order to meet the condition $\omega L/c=m\pi$, from which we can finally write:
\begin{equation}\label{eq:m}
m=\frac{2 L}{\lambda},
\end{equation} 
where $\lambda$ is the wavelength of the mode. Then, finally the phase shift in flat spacetime is:
\begin{equation}\label{eq:thetaf}
\theta_f=2\pi\frac{L}{\lambda}; \,\,\frac{L}{\lambda}=1,2,3\dots.
\end{equation} 

Let us now consider that instead of a flat spacetime, the field is propagating along the radial direction of a spacetime containing a traversable massless wormhole -- e.g. the paradigmatic Ellis wormhole \cite{ellis,morristhorne2}. Thus, the spacetime geometry is given by
\begin{equation}
ds^2=-c^2 dt^2+\frac{1}{1-\frac{b_0^2}{r^2}}\,dr^2, \label{eq:metric}
\end{equation}
where we are not considering the angular part of the metric.
There is a singular point of $r$ at which $r=b_0$, which determines the radius of the wormhole's throat and defines two different Universes or two asymptotically flat regions within the same Universe -as $r$ goes from $\infty$ to $b_0$ and then back from $b_0$ to $\infty$.  

\section*{Results}

In the wormhole spacetime, a free-falling observer possesses a spatial coordinate such that $l=\pm \sqrt{r^2-b_0^2}$. In the coordinates $l, t$ light will describe the trajectory $| l_2-l_1|= c\,t$. Therefore, in the laboratory coordinates we will have that the propagation length is different:
\begin{equation}
L'=\left|\sqrt{r_2^2-b_0^2}-\sqrt{r_1^2-b_0^2}\right|.
\end{equation}
The wavelength $\lambda$ of the mode will be modified as well. 
Let us consider that the propagation takes place between two points $r_1$, $r_2$ in the same branch of the Universe and very far from the wormhole throat $r_1, r_2 >>b_0$. We assume that the separation $L$ is small as compared to either point $L<<r1,r2$ and for simplicity we consider that $r2>r1$, thus $L=r_2-r_1$. Finally, the wavelength $\lambda$ is negligible with respect to the separation $L$, $L>>\lambda$. Under these approximations and taking into account Eq. (\ref{eq:thetaf}) the phase shift in the wormhole spacetime is 
\begin{equation}\label{eq:thetawh}
\theta=\theta_f\left(1- \frac{b_0^2}{2 r_1^2}\frac{L}{r_1}\right),
\end{equation}
where we see that the second term in the parenthesis represents a small correction to the flat spacetime phase shift $\theta_f$.

Notice that under all the above approximations Eq. (\ref{eq:metric}) reduces to a quasiflat spacetime given by the metric:
\begin{equation}\label{eq:effmet}
g_{\mu\nu}=\eta_{\mu\nu}+\widehat{g_{\mu\nu}}
\end{equation}
where $\eta_{\mu\nu}$ is the Minkowski metric and we only have a very small correction $\widehat{g_{\mu\nu}}$, where the only non-zero element is  $\widehat{g_{rr}}=b_0^2/r_1^2 <<1$. In other words, we are considering that the propagation of the coherent state takes place very far away from the wormhole, so that the spacetime is  almost completely flat and the effect of the wormhole  would be just a tiny perturbation. In this sense, the scenario resembles gravitational wave detection, where a different small perturbation of Minkowski spacetime is detected.

In order to find out whether this small correction of the phase shift is in principle detectable, we use that, by construction, $H(\theta)$ --or equivalently, $F(\theta)$ of the homodyne measurement scheme, which is equal to the QFI for the values of the phase that we are discussing, as mentioned above-- can be related to $H(b_0)$ through $H(b_0)=\left|\frac{\partial \theta}{\partial b_0}\right|^2 H(\theta)$, since both $H(x)$ and $F(x)$ are obtained through partial derivatives with respect to $x$ \cite{braunsteincaves} Then, using Eqs. (\ref{eq:qcr}), (\ref{eq:qfish}) and (\ref{eq:thetawh}) we find the following expression for the sensitivity relative to the value of $b_0$:
\begin{equation}\label{eq:sensrel}
\frac{\Delta b_0}{b_0}=\frac{\lambda}{4\pi L}\frac{r_1^2}{b_0^2}\frac{r_1}{L}\frac{1}{\sqrt{\langle n\rangle}}.
\end{equation}
If instead of using the QFI, we use the FI in Eq. (\ref{eq:fish}) we obtain:
\begin{equation}\label{eq:sensrelfi}
\frac{\Delta b_0}{b_0}=\frac{\lambda}{4\pi L}\frac{r_1^2}{b_0^2}\frac{r_1}{L}\frac{1}{\sqrt{\langle n\rangle\cos{\theta}}},
\end{equation}
where $\theta$ is given by Eq. (\ref{eq:thetawh}) and thus $\cos{\theta}$ is very close to 1.

We can consider a laser of $\lambda=10^3 \operatorname{nm}$ and a few $\operatorname{W}$ of power, which amounts to an average number of photons of $\langle n\rangle\simeq 10^{22}$ per second, as in LIGO \cite{rafalligo}. Using the average number of photons per second in Eqs. (\ref{eq:sensrel}) and (\ref{eq:sensrelfi}) gives rise to figures of merit with units of $\operatorname{Hz}^{-1/2}$, as is standard in laser interferometry. For the length $L$, we can consider the arm length of long-baseline inteferometers, ranging from a few $\operatorname{km}$ --LIGO-- to a few million $\operatorname{km}$ --LISA.  The order of magnitude of $b_0$ is unknown and the only limitation in the current approach is the sensible approximation $b_0<<r_1$, which implies that we stay in a quasiflat spacetime sufficiently far from the wormhole throat. 

\begin{figure}[h!] 
\includegraphics[width=0.95\linewidth]{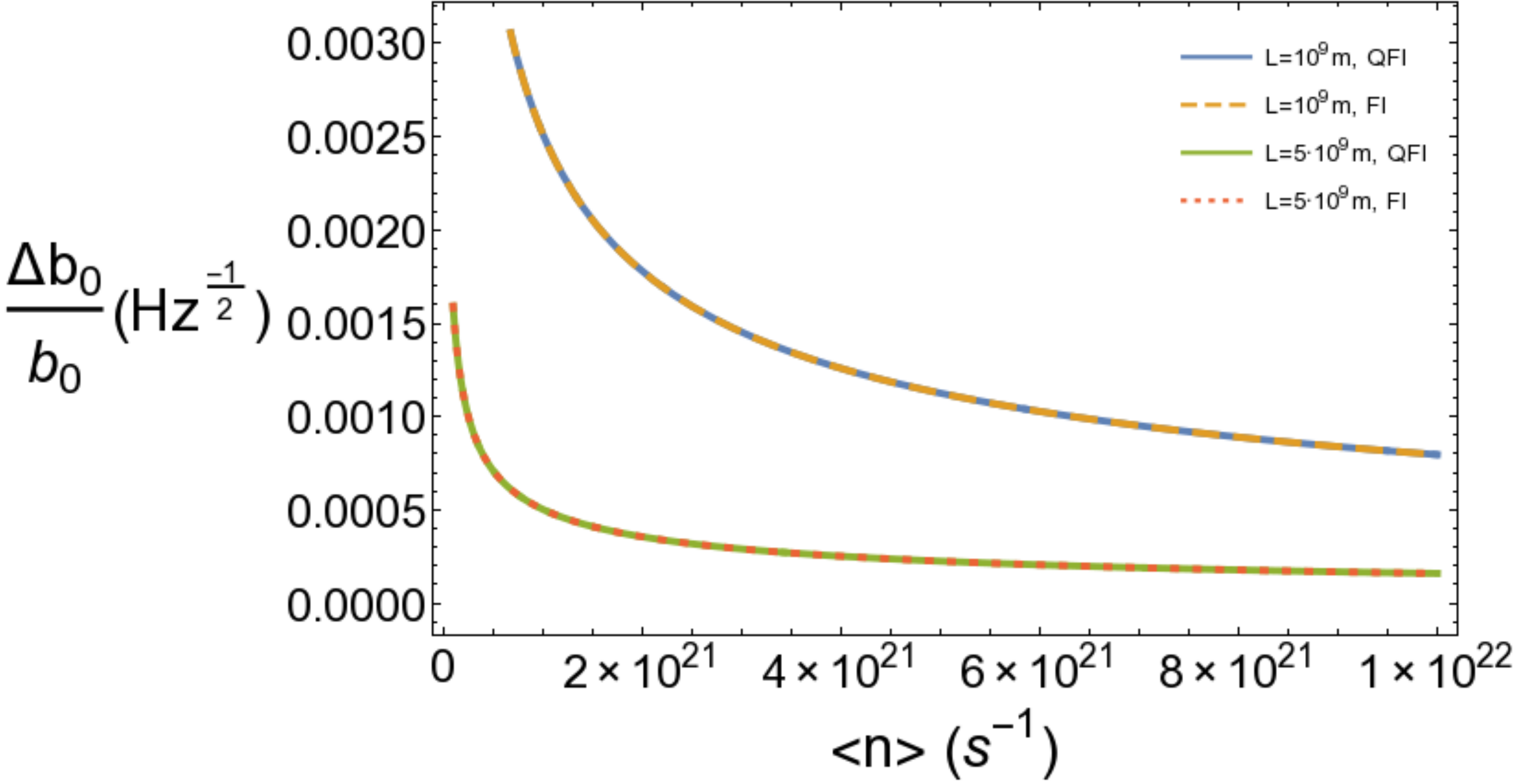}
 \caption{Figure of merit of the sensitivity of our scheme vs. average number of photons per second for $\lambda=10^{3}\operatorname{nm}$, $r_1/b_0=10^{11}$, $r_1/L=10^2$ and two different values of $L$ in the range of millions of $\operatorname{km}$. In both cases we plot both the optimal bound provided by the QFI and the FI of homodyne detection, which turn out to be practically indistinguishable.}
\label{fig2}
\end{figure}

In Fig. \ref{fig2} we see that in the regime of lengths of the order of LISA's interferometer arms, the QFI and the FI are practically indistinguishable and we are able to achieve a sensitivity which is significantly smaller than $b_0$ for distances $r_1= 10^{11}b_0$ and $L=10^9 b_0$. Note that for these values of $L$, then $b_0$ is of the order of $m$, and $r_1$ is of the order of $10 \operatorname{\mu pc}$. The detection by gravitational lensing effects requires a much larger value of the throat radius $0.1-10^5 \operatorname{pc}$ \cite{search}. Considering this range of values, our Fig. (\ref{fig2}) would correspond to $r_1=10^{5}- 10^{9} \operatorname{pc}$. Note that the estimated distance between the Earth and the rotational center of the Milky Way is around $10^4 \operatorname{pc}$ \cite{estimates}. Indeed, the sensitivity would be much better if we consider smaller values of $r_1/b_0$, always respecting $r_1>>b_0, L$ and taking into account that it does not seem realistic to assume that $r_1$ is close to the detection point. If we consider that the maximum error that we want to tolerate corresponds to $\Delta b_0/b_0=0.1 \operatorname{Hz}^{-1/2}$, then with the values of Fig. \ref{fig2} we can tolerate distances up to $r_1/b_0= 10^{12}$ with $r_1=10^2 L$ which amounts to a minimum throat radius of $b_0\simeq 10\operatorname{cm}$. These values for $b_0$ are significantly smaller than the ones detectable by gravitational lensing. 

\begin{figure}[h!] 
\includegraphics[width=0.95\linewidth]{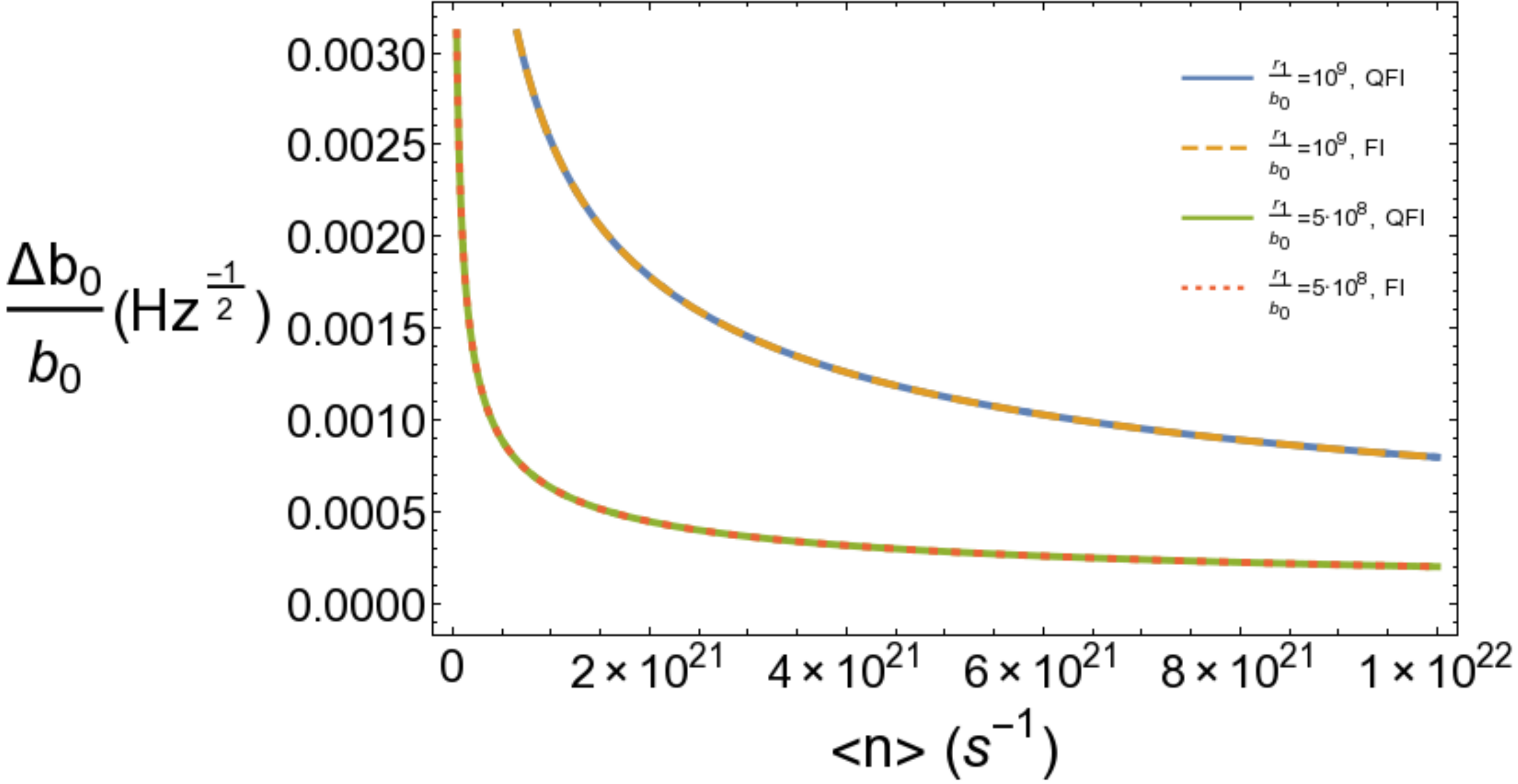}
 \caption{Figure of merit of the sensitivity of our scheme vs. average number of photons per second for $\lambda=10^{3}\operatorname{nm}$, $r_1/b_0=10^{5}$, $r_1/L=10^8$ and two different values of $L$ in the range of $\operatorname{km}$.}
\label{fig3}
\end{figure}

In Fig. (\ref{fig3}) we consider much smaller lengths $L\simeq10^3 \operatorname{m}$, as the interferometer arms in LIGO. This amounts to a restriction of the maximum allowed values of the parameter $r_1/b_0$, if we keep the same values for $r_1$ -- thus $r_1/L$ is larger than in Fig. (\ref{fig2}). We consider $10^5$ in the figure and we could go up to $10^{6}$ given the threshold for the error established above. This increases the minimum values of the throat radius that we could consider for a fixed large distance, as compared to the scenario in Fig. (\ref{fig2}). 

\section*{Discussion}

\begin{figure}[t!] 
\includegraphics[width=0.95\linewidth]{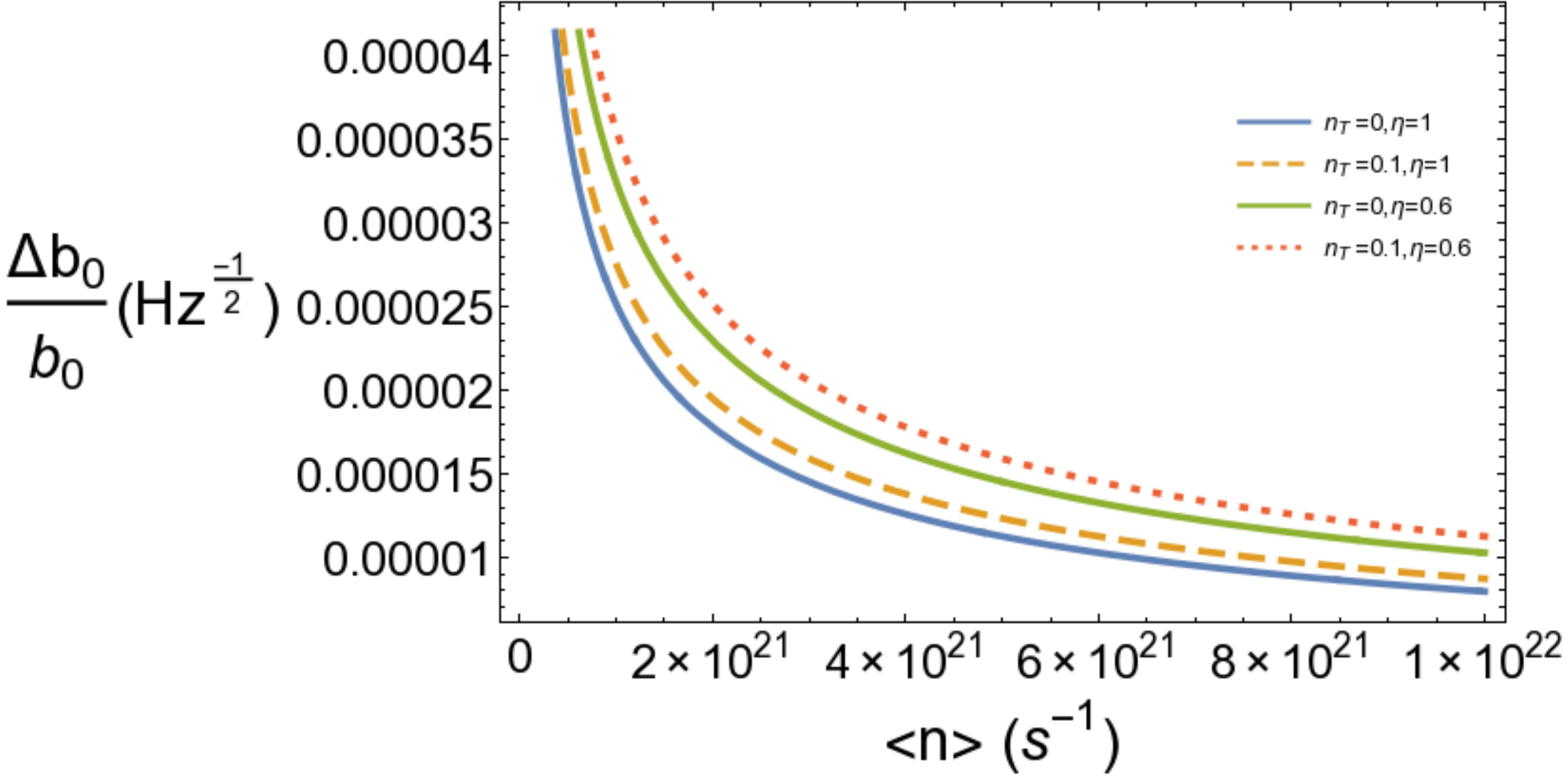}
 \caption{Figure of merit of the sensitivity of our scheme vs. average number of photons per second in the presence of optical losses and initial mixedness for $\lambda=10^{3}\operatorname{nm}$, $r_1/b_0=10^{7}$, $r_1/L=10^5$ and $L= 10^6\operatorname{m}$. We see that the combined effect of optical loss and mixedness has a moderate impact on the sensitivity.}
\label{fig4}
\end{figure}

Let us discuss the effect of some possible deviations from the theoretical scenario described so far. Since we are within the standard scheme of phase estimation with coherent states and homodyne detectors, we can take advantage of some results in the literature. For instance, the presence of optical losses would affect the sensitivity by multiplying the FI by a factor  $\eta$ \cite{losses}, where $\eta$ is a number between 0 and 1 accounting for the optical efficiency. In modern laser interferometers the value of $\eta=0.62$ \cite{rafalligo} is achieved, which implies that optical losses are $38 \%$.  We can also consider that the initial state is not the pure state resulting from the action of the displacement operator onto the vacuum, but onto a mixed state characterized by a thermal distribution with average number of excitations $n_T$. Although for the optical frequencies that we are considering the actual number of thermal photons would be negligible even at room temperature on Earth, it can be useful to consider a relatively high $n_T$ in order to see the impact of any possible initial mixedness. The FI in this case would be reduced by a factor $(1+2n_T)^{-1}$ \cite{aspachs}. Putting all together, the expected sensitivity in the presence of optical losses and initial mixedness would be given by:
\begin{equation}\label{eq:sensrelfietaT}
\frac{\Delta b_0}{b_0}=\frac{\eta}{1+2n_T}\frac{\lambda}{4\pi L}\frac{r_1^2}{b_0^2}\frac{r_1}{L}\frac{1}{\sqrt{\langle n\rangle\cos{\theta}}}.
\end{equation}
In Fig. (\ref{fig4}) we see the robustness of our scheme in the presence of realistic optical losses and relatively high initial mixedness. 
Further analysis of error sources would strongly depend on the particular details of the experimental setup and lies beyond the scope of this work. We can anticipate however, that an eventual experimental test would be of course extremely challenging, e.g. an extremely careful calibration of any gravity gradient noise would be required, as in LIGO. However, notice that LIGO has been able to detect a phase sift of approximately 10-10 rad. Using Eqs. (9) and (12) and the values of $L$ and $\lambda$ considered in this work, we find that in our case this means $\pi\frac{b_0^2}{r_1^3}\simeq 10^{-22}\operatorname{m}^{-1}$ for $L= 1\operatorname{km}$ and $\pi\frac{b_0^2}{r_1^3}\simeq 10^{-34}\operatorname{m}^{-1}$ for $L= 10^6\operatorname{km}$. While, in principle, there is no restriction on the value of $b_0$ ,we can focus on a case of interest, such as a ''black-hole mimicker'' \cite{gravastar,konoplya}. Then $b_0$ would be slightly larger than the Schwarzschild radius of the black hole, which in the case of the first LIGO detection would be around $200\operatorname{km}$. Putting all the numbers together, we find that detection would be possible at a distance ranging from $r_1\simeq10\operatorname{\mu Pc}$ (for $L=1\operatorname{km}$) to $r_1\simeq0.1\operatorname{Pc}$ (for $L=10^6\operatorname{km}$). Along these lines, it might be useful to recall that a quantum simulator of a wormhole spacetime for the electromagnetic field in the $\operatorname{GHz}$ regime has been recently proposed \cite{qsgw}, which could be an important low-cost Earth-based source of information for the actual --presumably space-borne-- experiments.

In summary, we have shown that it is, in principle, possible to detect a distant wormhole in our spacetime by measuring the correction to the phase shift of an electromagnetic field propagating over large distances. We have considered a standard scenario with a single-mode coherent state with a large average number of photons and homodyne detection, as well as state-of-the-art numbers in modern laser interferometry. The relevant parameters turn out to be the ratio between the radius of the wormhole throat $b_0$ and the distance to the wormhole $r_1$, together with the ratio between the propagation distance $L$ and $r_1$. While these ratios cannot be large --since $r_1>>b_0, L$-- we take advantage of the large average number of photons per second and the large value of $L/\lambda$. We show that detection is in principle possible for $r_1/b_0$ up to $10^{6} - 10^{12}$, which allows us to consider a wide range of throat radii and distances. The scheme is in principle highly robust to optical losses and initial mixedness. While of course we are aware that an actual experimental test based on these ideas would be extremely challenging, we hope that the results of this work are promising enough to open a new avenue of research on the detection of one of the most fascinating objects that might --or might not-- exist in Nature.

\section*{Acknowledgements.}  Financial support by Fundaci{\'o}n General CSIC (Programa ComFuturo) is acknowledged by C.S as well as additional support from Spanish MINECO/FEDER FIS2015-70856-P and CAM PRICYT Project QUITEMAD+ S2013/ICE-2801.

\section*{Additional information} 
\subsection*{Competing financial interests}
The author declares no competing financial interests.

\end{document}